\documentclass[twocolumn]{aastex631}
\usepackage{amsmath,amstext}
\usepackage{hyperref}
\usepackage{amssymb}
\usepackage{wasysym}
\usepackage{float}
\usepackage{ulem}
\usepackage{breakurl}
\usepackage{color}

	\newcommand{\msun}{M_{\odot}}
    
    \newcommand{\fescH}{\bar f_{\rm esc,H}}
    \newcommand{\fescHe}{\bar f_{\rm esc,He}}

\def\spose#1{\hbox to 0pt{#1\hss}}
\def\lta{\mathrel{\spose{\lower 3pt\hbox{$\mathchar"218$}}
     \raise 2.0pt\hbox{$\mathchar"13C$}}}
\def\gta{\mathrel{\spose{\lower 3pt\hbox{$\mathchar"218$}}
     \raise 2.0pt\hbox{$\mathchar"13E$}}}
     
\def\HI{\hbox{\rm H\,$\scriptstyle\rm I$}}
\def\H{\hbox{\rm H}}
\def\HII{\hbox{\rm H\,$\scriptstyle\rm II$}}
\def\HeI{\hbox{\rm He\,$\scriptstyle\rm I$}}

\def\HeII{\hbox{\rm He\,$\scriptstyle\rm II$}}
\def\HeIII{\hbox{\rm He\,$\scriptstyle\rm III$}}
\newcommand{\QH}{Q_{\mathrm{HII}}}
\newcommand{\QHe}{Q_{\mathrm{HeIII}}}
\def\emiss{\,{\rm erg\,s^{-1}\,cMpc^{-3}\,Hz^{-1}}}

\setwatermarkfontsize{6cm}

\righthead{AGN-Dominated Reionization}
\lefthead{Madau et al.}

\journalinfo{Submitted to AAS Journals}

\begin{document}

\title{Cosmic Reionization in the JWST Era: Back to AGNs?}

\author[0000-0002-6336-3293]{Piero Madau}
\affiliation{Department of Astronomy \& Astrophysics, University of California, 1156 High Street, Santa Cruz, CA 95064, USA}
\affiliation{Dipartimento di Fisica ``G. Occhialini", Universit\`a degli Studi di Milano-Bicocca, P.za della Scienza 3, I-20126 Milano, Italy}

\author{Emanuele Giallongo}
\affiliation{INAF-Osservatorio Astronomico di Roma, Via Frascati 33, I-00078, Monte Porzio Catone, Italy}

\author{Andrea Grazian}
\affiliation{INAF-Osservatorio Astronomico di Padova, Vicolo dell'Osservatorio 5, I-35122, Padova, Italy}

\author{Francesco Haardt}
\affiliation{Dipartimento di Scienza e Alta Tecnologia, Universit\`a degli Studi dell'Insubria, via Valleggio 11, I-22100 Como, Italy}
\affiliation{INAF, Osservatorio Astronomico di Brera, Via E. Bianchi 46, I-23807 Merate, Italy}
\affiliation{INFN, Sezione Milano-Bicocca, P.za della Scienza 3, I-20126 Milano, Italy}

\begin{abstract}
Deep surveys with the James Webb Space Telescope (JWST) have revealed an emergent population of moderate-luminosity, broad-line active galactic nuclei (AGNs) at $4\lta z\lta 13$ powered by accretion onto early massive  black holes. The high number densities reported, together with the large Lyman-continuum (LyC) production efficiency and leakiness into the intergalactic medium that are typical of UV-selected AGNs, lead us to reassess a scenario where AGNs are the sole drivers of the cosmic hydrogen/helium reionization process. Our approach is based on the assumptions, grounded in recent observations, that: (a) the fraction of broad-line AGNs among galaxies is around $10-15\%$; (b) the mean escape fraction of hydrogen LyC radiation is high, $\gta 80$\%, in AGN hosts and is negligible otherwise; and (c) internal absorption at 4 ryd or a steep ionizing EUV spectrum delay full reionization of \HeII\ until $z\simeq 2.8-3.0$, in agreement with observations of the \HeII\ Lyman-$\alpha$ forest.
In our fiducial models, (1) hydrogen reionization is 99\% completed by redshift $z\simeq 5.3-5.5$, and reaches its midpoint at $z\simeq 6.5-6.7$; (2) the integrated Thomson scattering optical depth to reionization is $\simeq 0.05$, consistent with constraints from cosmic microwave background anisotropy data; and (3) the abundant AGN population detected by JWST does not violate constraints on the unresolved X-ray background. 
\end{abstract}


\section{Introduction}

The reionization of intergalactic hydrogen during the first billion years of cosmic history is intimately tied to the physics of structure formation, the thermodynamics of diffuse baryonic matter in the Universe, and the nature of the first astrophysical sources of radiation and heating \citep[see][for a recent review]{Gnedin2022}. Recent observations with the James Webb Space Telescope (JWST) have enabled the spectroscopic confirmation of infant, low-luminosity galaxies at $z\simeq 6-14$ \citep[e.g.,][]{Curtis-Lake2023, FujimotoCEERS2023,Wang2023,Roberts-Borsani2023,Robertson2023a,Carniani2024} and the determination of the neutral hydrogen fraction and radii of ionized bubbles during the epoch of reionization \citep[e.g.,][]{Umeda2023,Heintz2023}, and they may be providing new clues and constraints on the contribution of different sources to the ionizing photon budget of the Universe \citep[e.g.,][]{Mascia2024,Atek2024,Endsley2023,Simmonds2024,Munoz2024}.

It is the conventional view that, as the declining population of optically bright quasars makes an increasingly small input to the 1 ryd EUV background at $z>4$, massive stars in young galaxies take over and provide the additional hydrogen-ionizing photons required at early times \citep[e.g.,][and references therein]{haardt12, Gnedin2014b,puchwein19,Faucher2020,Dayal2020, Yeh2023,Atek2024}. To date, however, only a few reliable star-forming galaxies at high redshifts are known to be copious Lyman-continuum (LyC) leakers \citep[e.g.,][]{Steidel2018,Vanzella2018,Fletcher2019,Rivera2019}, arguably leaving much of the radiation necessary to reionize the Universe still unaccounted for. For instance, from stacked images of 165 faint star-forming galaxies in the Hubble Ultra-Deep Field combined with deep Multi-Unit Spectroscopic Explorer observations, \citet{Japelj2017} estimated a $1\sigma$ upper limit to the mean escape fraction of $\fescH <0.07$ in the $3<z<4$ redshift range. In a UVCANDELS sample of 90 star-forming galaxies with secure redshifts in the range $2.4-3.7$, only 5 were identified as potential LyC leakers, while most stacks of the 85 nondetection galaxies gave tight 2$\sigma$ upper limits below $\fescH <0.06$ \citep{Wang2023fesc}. Similarly, no LyC leakage ($\fescH<0.063\pm 0.007$) has been detected by \citet{Naidu2018} in a stack of extreme [OIII] emitters (a ``Green Pea" subgroup known to have an elevated $\fescH$ at low redshifts; e.g., \citealt{Izotov2018}) or in low-mass lensed galaxies at $1.3\le z\le 3.0$ \citep{Jung2024}. {In a sample of 621 Ly$\alpha$-emitting galaxies at $z=3.0-4.5$, \citet{Kerutt2024} recently identified only 5 likely LyC leakers but argued that this detection rate was consistent with a global escape fraction of 12\%.} Using an alternative approach to constrain $\fescH$ empirically at $1.6<z<6.7$ via spectroscopy of long-duration gamma-ray burst afterglows, \citet{Tanvir2019} estimated an average escape fraction of 0.7\% at the Lyman limit.

Active galactic nuclei (AGNs), on the other hand, are known to leak a large fraction of their hydrogen LyC radiation into the intergalactic medium (IGM). No discernible continuum edge at 1 ryd  has been detected in UV-bright, $z<1.5$ quasars \citep[$\bar \tau_{912}<0.01$,][]{Stevans2014}. In a large sample of bright Sloan Digital Sky Survey quasars at $3.6\le z\le 4.6$,\footnote{At $z\gta 6$, owing to the opacity of the IGM, the LyC emission cannot be measured directly.}\ \citet{Romano2019} found a lower limit to the mean escape fraction of $\fescH>0.49$. This measurement is in agreement with the values obtained for fainter ($\fescH=0.44-1$; \citealt{Grazian2018}) and brighter AGNs  ($\fescH=0.75$,  \citealt{Cristiani2016}) at similar redshifts.\footnote{Note that all these values are a measurement of the relative escape fraction, the ratio of fractions of leaking ionizing to nonionizing UV photons, as defined by the cited authors.}
{Somewhat lower leakages ($\sim 0.30-0.5$) have been inferred by \citet{Micheva2017} and \citet{Iwata2022}. Their procedures are based on AGN broadband colors, however, and
the derived $\fescH$ values are affected by assumptions on the intrinsic AGN spectra and IGM mean free paths along the specific lines of sight.} AGNs are also characterized by very high LyC production rates. In a study of 111 massive star-forming galaxies with and without an AGN at $2.26 < z <4.3$, \citet{Smith2020} found that a stack of 17 AGNs dominated the LyC production by a factor of 10 compared to all 94 galaxies without an AGN. We observe that a significant number of AGN candidates have been recently identified in Green Pea galaxies using mid-IR and X-ray observations \citep{Harish2023,Singha2024}.

Deep surveys with the JWST have recently discovered an emergent population of moderate-luminosity type 1 AGNs at $4\lta z\lta 13$ powered by accretion onto early $M_{\rm BH}=10^{5}-10^8\,\msun$ black holes and revealed through the detection of their broad-line region (BLR) as seen in the Balmer emission lines \citep{Kocevski2023}. The AGN number fractions reported are very high, from $5-15\%$ at $z\sim 4-7$ \citep{Harikane2023AGN,Maiolino2023} to $>10-35\%$ at $z\ge 8.5$ \citep{FujimotoUNCOVER2023}. This is much higher than an extrapolation of the quasar luminosity function (LF) and implies a large, previously missing \citep[but see][]{giallongo15,giallongo2019} population of faint AGNs in the early Universe. The broad-line AGNs in the JADES sample of \citet{Maiolino2023} have H$\alpha$/H$\beta$ Balmer decrements generally consistent with the Baker-Menzel case B value of 2.8, which is indicative of little dust attenuation. A subcomponent of JWST-selected broad-line AGNs is compact, with red continua in the rest-frame optical wavebands (``little red dots") but with blue slopes in the UV \citep[e.g.,][]{Greene2024,Kokorev2024,Matthee2024,Perez-Gonzalez2024}. One such reddened AGN (ID20466 at $z=8.50$) seems to reside in a huge ionized bubble, perhaps indicating that the covering fraction of the dusty layers surrounding the AGN is minimal and the leakage of ionizing photons is high \citep{FujimotoUNCOVER2023}.

The surprising near-ubiquity of JWST-selected type 1 AGNs at $z>5$, together with the large LyC production efficiency and leakiness into the IGM that are typical of UV-selected AGNs, lead us to reassess a scenario where AGNs are the main drivers of the cosmic reionization process \citep[e.g.,][]{chardin15,giallongo15,madau15,giallongo2019}.  We explore this  possibility below.  All the calculations presented in this paper assume a flat cosmology with parameters $(\Omega_M, \Omega_\Lambda, \Omega_b) = (0.3, 0.7, 0.045)$, a Hubble constant of $H_0=70\,$ km s$^{-1}$ Mpc$^{-1}$, and a primordial baryonic gas of total hydrogen and helium mass fractions $X=0.75$ and $Y=0.25$. We shall denote with $h\nu_{912}$, $h\nu_{504}$, and $h\nu_{228}$ the ionization energies of \HI, \HeI, and \HeII, respectively. 
 
\section{AGN FUV Emissivity}

The basic premise of our model is that a highly ionizing minority of galaxies -- those hosting an AGN -- accounts for the bulk of the reionization photon budget of the Universe. It is generally understood that the escape of LyC radiation from the centers of these systems into the IGM is regulated by stellar \citep[e.g.,][]{Ma2020,Katz2023} and AGN feedback \citep{Trebitsch2021}. The dearth of LyC leakers among non-AGN star-forming galaxies may be indicative of a possible temporal bimodality in $f_{\rm esc,H}$, whereby inactive galaxies are opaque to LyC radiation, and  only during their AGN phase low-density channels -- through which ionizing photons can easily escape -- are opened in their interstellar medium (ISM). 

Interestingly, nearby LyC-emitting galaxies with resolved X-ray emission are observed to host variable, accretion-powered sources that are likely low-luminosity AGNs \citep{Kaaret2022}. The role of X-rays in boosting the escape fractions for both hydrogen and helium has been modeled by \citet{Benson2013}. Evidence for an AGN-aided pathway for the escape of ionizing radiation from candidate LyC leakers at $ z\sim 2$ is also seen in deep UV Hubble Space Telescope (HST) imaging \citep{Naidu2017}. 

\begin{figure}[!ht]
\centering
\includegraphics[width=\hsize]{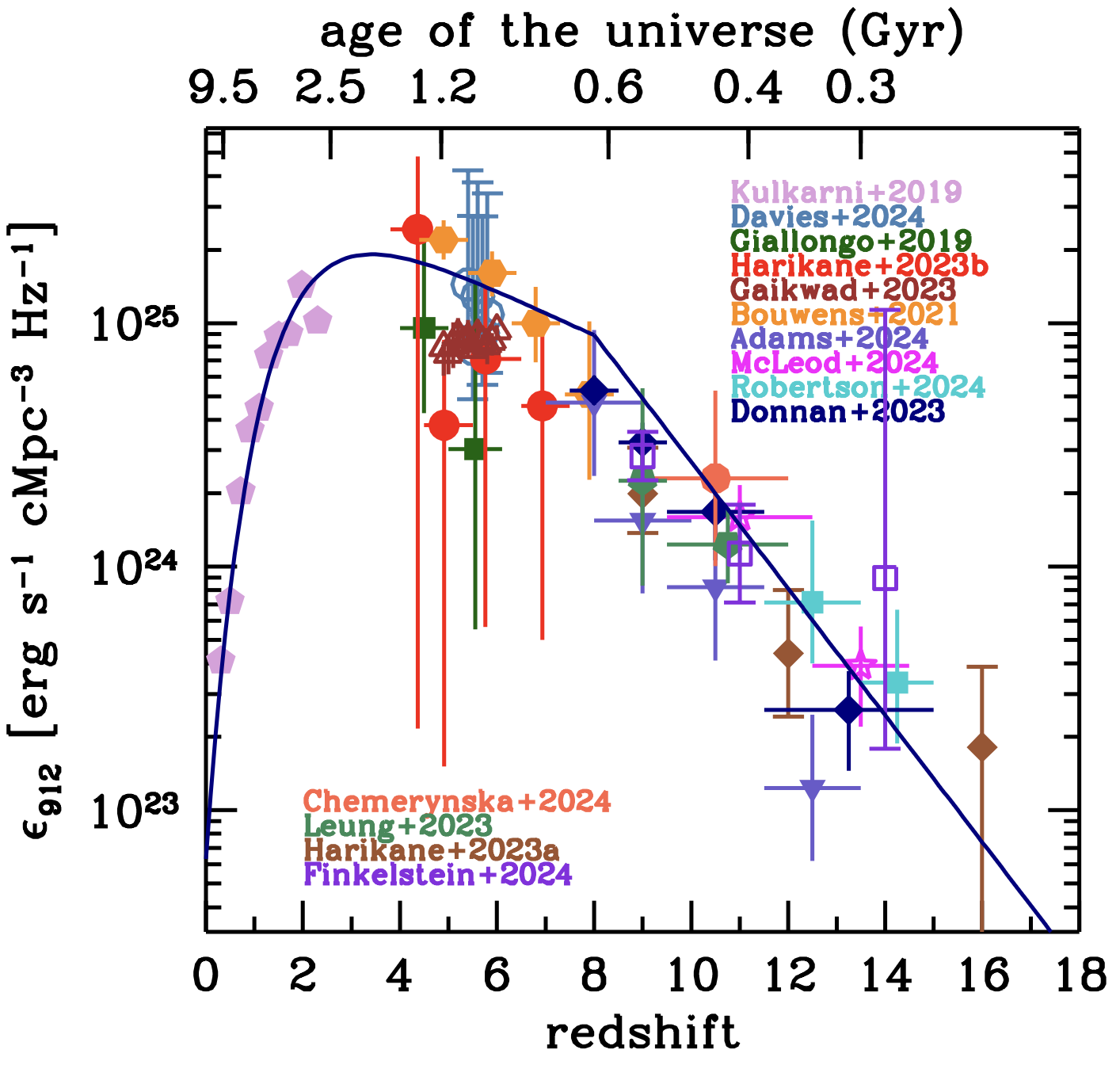}
\caption{The ``AGN$+$host" comoving emissivity {on the red side of the Lyman limit}. The $z\ge 5$ galaxy emissivities from 
\citet[][]{Bouwens2021} (orange exagons),
\citet{Adams2024} (slate-blue inverted triangles),
\citet{Leung2023} (green pentagons),
\citet{McLeod2024} (magenta stars),
\citet{Harikane2023GAL} (brown diamonds),
\citet{Donnan2023} (navy blue diamonds), 
\citet{Chemerynska2023} (dark-orange dot),
\citet{Finkelstein2023} (purple empty squares),
and \citet{Robertson2024} (cyan squares) have been multiplied by a type 1 AGN fraction of $f_{\rm AGN}=0.15$ following \citet{Maiolino2023} and 
\citet{FujimotoUNCOVER2023}. We have also 
plotted determinations of the AGN emissivity 
from \citet{Kulkarni2019} (purple pentagons),  \citet{giallongo2019} (green squares), and
\citet{Harikane2023AGN} (red dots). All high-$z$ LFs have been integrated down to $M_{\rm UV}=-17$ and we have converted the UV luminosity density from these studies into an emissivity at the Lyman limit using a power-law spectral slope of $\alpha_{\rm FUV}=-0.61$ \citep{Lusso2015}. The 
estimates of \citet{Gaikwad2023} (brown triangles) and \citet{Davies2024} (steel-blue
empty circles) from the statistics of the Lyman-$\alpha$ forest account for the LyC leakage into the IGM. The solid curve shows the functional form given in Equation (\ref{eq:emiss}).
}
\label{fig:em912}
\end{figure}

To assess whether type 1 AGNs can dominate the cosmic reionization process under reasonable physical assumptions, we show in Figure \ref{fig:em912} our adopted intrinsic comoving emissivity, $\epsilon_{912}(z)$, on the red side of the Lyman limit -- i.e. unaffected by the Lyman discontinuity in young stellar population spectra and uncorrected for partial source obscuration and leakage effects --  from type 1 AGNs$+$hosts. All the optical surveys cited below provide best-fit LF parameters, which have been used to integrate the LF down to an absolute magnitude $M_{\rm UV}=-17$ (the limit of the deepest JWST spectroscopic surveys). The steepness of the faint-end slope of the LF at high redshifts makes the integrated emissivity rather sensitive on this limiting magnitude. We have converted the nonionizing UV luminosity density from these studies into an  emissivity at the Lyman limit using a power-law spectral slope of $\alpha_{\rm FUV}=-0.61$ \citep{Lusso2015}. 

The determinations of the UV LF of broad-line AGNs all assume, following \citet{Harikane2023AGN}, 
\begin{equation}
\Phi_{\rm AGN}(M_{\rm UV},z)=f_{\rm AGN}(M_{\rm UV},z)\,\Phi(M_{\rm UV},z),
\end{equation}
where $f_{\rm AGN}$ is the AGN fraction in each magnitude and redshift bin, and the fitting parameters of the galaxy LF $\Phi$ are taken from \citet{Bouwens2021} ($5<z<8$),
\citet{Adams2024} ($8<z<12.5$), 
\citet{Chemerynska2023} ($z=10.5$)
\citet{Harikane2023GAL} ($9<z<16$),
\citet{Donnan2023} ($8<z<13.25$), 
\cite{Leung2023} ($9<z<11$),
\citet{McLeod2024} ($z=11,13.5$),
\citet{Finkelstein2023} ($9<z<14$),
and \citet{Robertson2024} ($12.5<z<14.25$). We assume $f_{\rm AGN}=0.15$ at all $z\ge 5$,
with no dependence on magnitude. We do not explicitly include in our tally the large, $f_{\rm AGN}=0.2\pm 0.03$, population of highly obscured, narrow-line type 2 AGNs at high redshift recently identified at $z=4-6$ with JWST spectroscopy \citep{Scholtz2023}. We have also added estimates of $\epsilon_{912}(z)$ from (1) broad-line AGNs (and their hosts) at $4.4<z<7$ by \citet{Harikane2023AGN}\footnote{Note that the AGN emissivities quoted in \citet{Harikane2023AGN} assume a 50\% escape fraction and subtract an estimated contribution of the host galaxy equal to 50\% in the rest-frame UV. For consistency with all other determinations, their values of $\epsilon_{912}$ should be scaled up by a factor of 4.}; (2) X-ray-selected AGNs at $z=4.5$ and $5.55$ \citep{giallongo2019}; (3) comparing the statistics of the Lyman-$\alpha$ forest at $4.9<z<6.0$ with cosmological simulations \citep[these emissivities account for the escape fraction into the IGM,][]{Gaikwad2023,Davies2024}; and (4) a low-redshift sample of color-selected AGNs  for illustration \citep{Kulkarni2019}. Despite some significant scatter, the function 
\begin{equation}
{\cal E}(z)=
\begin{cases}
25.7e^{-0.0037z}-2.9e^{-z}&~~~(z\le 8);\\
{\cal E}(8)-0.26(z-8)&~~~(z>8),
\end{cases}
\label{eq:emiss}
\end{equation}
where ${\cal E}\equiv \log_{10}\epsilon_{{912}}$ and
$\epsilon_{912}$ is expressed in units of $\emiss$, reproduces the data points reasonably well (solid line in Figure \ref{fig:em912}). Note that this AGN emissivity is more peaked and drops faster at high redshifts than the one adopted in \citet{madau15}.

\section{Photon Source and Sink Terms}

\begin{figure}[!hb]
\centering
\includegraphics[width=\hsize]{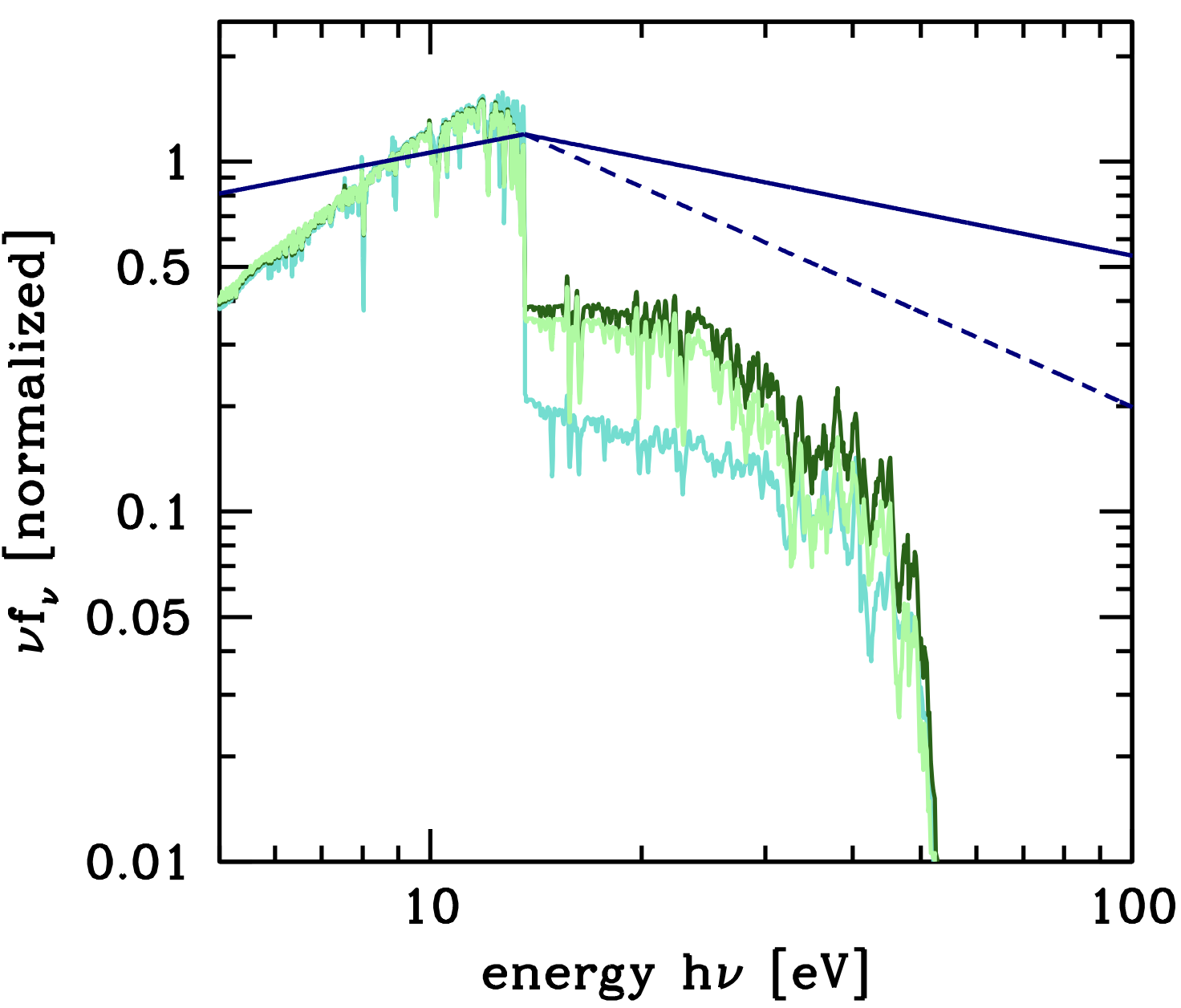}
\caption{The intrinsic (uncorrected for partial source obscuration and leakage effects) ionizing spectrum ($\nu f_\nu$) of a model type 1 AGN$+$host. The stellar and AGN contributions have been normalized to unity at wavelength $\lambda=1450\,$\AA. The AGN spectral template (blue curve) assumes a power-law $f_\nu\propto \nu^{-0.61}$ redward of the Lyman limit and $f_\nu\propto \nu^{\alpha_{\rm EUV}}$ at energies $>1$ ryd, with 
$\alpha_{\rm EUV}=-1.4$ (solid line) and $-1.9$ 
(dashed line). The green curves show BPASS spectral templates \citep{Eldridge2017} for galaxies with metallicity $Z=0.002$, the `imf135\_100' initial mass function, and no dust attenuation. Dark green: constant star formation at age 100 Myr; binary evolution model. Light green: constant star formation at age 100 Myr; single star evolution model. Turquoise: simple stellar population at age 10 Myr; binary evolution model.
}
\label{fig:bpass}
\end{figure}

Reionization is achieved when ionizing sources have radiated away at least one LyC photon per atom in the IGM, and the rate of LyC photon production is sufficient to balance radiative recombinations in a clumpy medium. Below, we shall denote with $\dot n_{\rm ion}$ the volume-averaged {\it injection rate} of LyC radiation into the IGM by type 1 AGNs. This is an integral over the photon spectrum between 1 and 4 ryd in the case of \HI\ ($\dot n_{\rm ion,H}$), and above 4 ryd for \HeII\ ($\dot n_{\rm ion,He}$):

\begin{equation}
\dot n_{\rm ion,H}(z)=\fescH
\int_{\nu_{912}}^{\nu_{228}}
{\epsilon_\nu(z)\over h\nu}d\nu,
\label{eq:ndotH}
\end{equation}
\begin{equation}
\dot n_{\rm ion,He}(z)=\fescHe
\int_{\nu_{228}}^{\infty}
{\epsilon_\nu(z)\over h\nu}d\nu.
\label{eq:ndotHe}
\end{equation}
The quantities $\fescH$ and $\fescHe$ in the above equations are some suitable averages over frequency and source luminosity of the absolute escape fraction of ionizing radiation relative to the dust-corrected observed UV flux.\footnote{Specifically, at a given redshift, we write the H-ionizing photon injection rate as
\begin{equation}
\begin{aligned}
\dot n_{\rm ion,H}= 
& \int\int_{\nu_{912}}^{\nu_{228}} ({L_\nu \over h\nu}) \phi_{\rm AGN}(L_\nu) f_{\rm esc,H}(\nu, L_\nu)\,dL_\nu d\nu\\
=& \int\int_{\nu_{912}}^{\nu_{228}} ({L_\nu \over h\nu}) \phi_{\rm AGN}(L_\nu) e^{-\tau_\nu(L_\nu)}\,dL_\nu d\nu\\
\equiv & \,\fescH\int_{\nu_{912}}^{\nu_{228}} {\epsilon_\nu\over h\nu}\,d\nu, 
\end{aligned}
\label{eq:fesc}
\end{equation}
and similarly for helium. Here, $\phi_{\rm AGN}(L_\nu)$ is the type 1 AGN LF, $e^{-\tau_\nu}$ is the frequency- and luminosity-dependent transmission through the nuclear region and the ISM of the host (averaged over all directions), and $\tau_\nu(L_\nu)$ is the optical depth to photoionization and dust absorption of a purely absorbing screen.}\, Note that we include in the escape parameters also the effect of partially absorbing material within the gravitational sphere of influence of the central black hole (see Section 3.1).
The AGN intrinsic ionizing emissivity can then be written as 
\begin{equation}
\epsilon_\nu= (1-\bar f_{\rm host})\,\epsilon_{{912}} \,(\nu/\nu_{912})^{\alpha_{\rm EUV}},
\label{eq:epsi}
\end{equation}
where $\bar f_{\rm host}$ is the population-averaged fractional input of the host galaxy to the rest-frame UV light, and we have adopted a single power law from 1 ryd to the soft X-ray band. In broad-line faint AGNs, the decomposition analysis of \citet{Harikane2023AGN} yields
$\bar f_{\rm host}\sim 50\%$. For $\bar f_{\rm host}\lta 0.5$, the stellar contribution to the ionizing emissivity can be neglected compared to the AGN nonthermal continuum to a first approximation. This is because of the intrinsic Lyman discontinuity in the spectra of massive stars \citep[e.g.,][]{Schaerer1997}, which causes a jump of a factor of 3 or more at 1 ryd even in the case of young, low-metallicity, dust-free stellar populations that include binary evolution pathways \citep[][see Figure \ref{fig:bpass}]{Eldridge2017}.

HST FUV observations yield for the composite rest-frame continuum of $z<1.5$ AGNs a slope of $\langle \alpha_{\rm EUV}\rangle =-1.41\pm 0.15$ \citep[$\lambda=500-1000\,$\AA,][]{Stevans2014}. A steeper slope, $\langle \alpha_{\rm EUV}\rangle =-1.70\pm 0.61$ ($\lambda=600-912\,$\AA), was measured at $z\simeq 2.4$ by \citet{Lusso2015},
while \citet{Zheng1997} estimated $\langle \alpha_{\rm EUV}\rangle =-1.96\pm 0.15$ ($\lambda=350-1050\,$\AA) in a sample of low-redshift, radio-loud and radio-quiet quasars. The mean quasar spectral energy distribution (SED) is poorly constrained at $\lambda<500\,$\AA, which impacts the \HeII\ photoionization rate 
($\lambda<228\,$\AA), and makes models of the helium-ionizing background highly sensitive to the assumptions. Using opacity measurements of the \HeII\ Lyman-$\alpha$ forest at $2.5< z< 3.2$ from \citet{worseck16}, \citet{Khaire2017} concluded that the slope of the AGN SED at energies $\ge 4$ ryd must lie in the range $-2<\alpha_{\rm EUV}<-1.6$, with a preferred value of $-1.8$. In general, observations of \HeII/\HI\ ratios in QSO absorbers suggest either an ionizing SED considerably softer than $-1.5$ or a partial flux blockage in the emergent 4 ryd continuum, with $\fescHe\approx 0.5-0.9$ \citep{Shull2020}. 

\subsection{LyC Leakage}

In contrast to the case of hydrogen, there are little data and few detailed theoretical studies \citep[but see][]{Benson2013} of the escape fraction of radiation capable of doubly ionizing helium, $\fescHe$, from AGNs. Because of the high ionization threshold and small photoionization cross-section of \HeII, and the rapid recombination rate of \HeIII, one expects $\fescHe\le \fescH$. Indeed, in a recent study of internal AGN absorption at 4 ryd, \citet{Shull2020} observed that the measured $\HeII/\HI$ absorption ratios in the Lyman-$\alpha$ forest at $2.4<z<2.9$ require $\fescHe/\fescH=0.6-0.8$. Internal AGN absorption may also be responsible for the small \HeII\ proximity regions seen in some AGNs \citep{Schmidt2018}.

A realistic scenario where EUV radiation from type 1 AGNs dominate the reionization budget may have to account for the effect of partially absorbing material in a small region enclosed within the gravitational sphere of influence of the central black hole. Intrinsic obscuration does indeed play a fundamental role for our understanding of the overall properties of AGNs \citep[e.g.,][]{Elitzur2008,Merloni2014,Ricci2017, Vito2018}. For instance, more than 60\% (and perhaps as many as 90\%) of the observed type 1 AGNs are known to present signatures of moderately ionized gas in their soft X-ray spectra, the so-called warm absorber phenomenon \citep[e.g.,][]{Reynolds1995,Blustin2005, Laha2014}. Located at pc-scale distances, and with absorbing hydrogen columns in the range $N_{\rm H}=10^{20-23}\,$cm$^{-2}$ and high ionization parameters, such warm absorbing clouds may moderate the leakage of helium-ionizing $>4$ ryd photons from AGN hosts. 

\begin{figure}[!ht]
\centering
\includegraphics[width=\hsize]{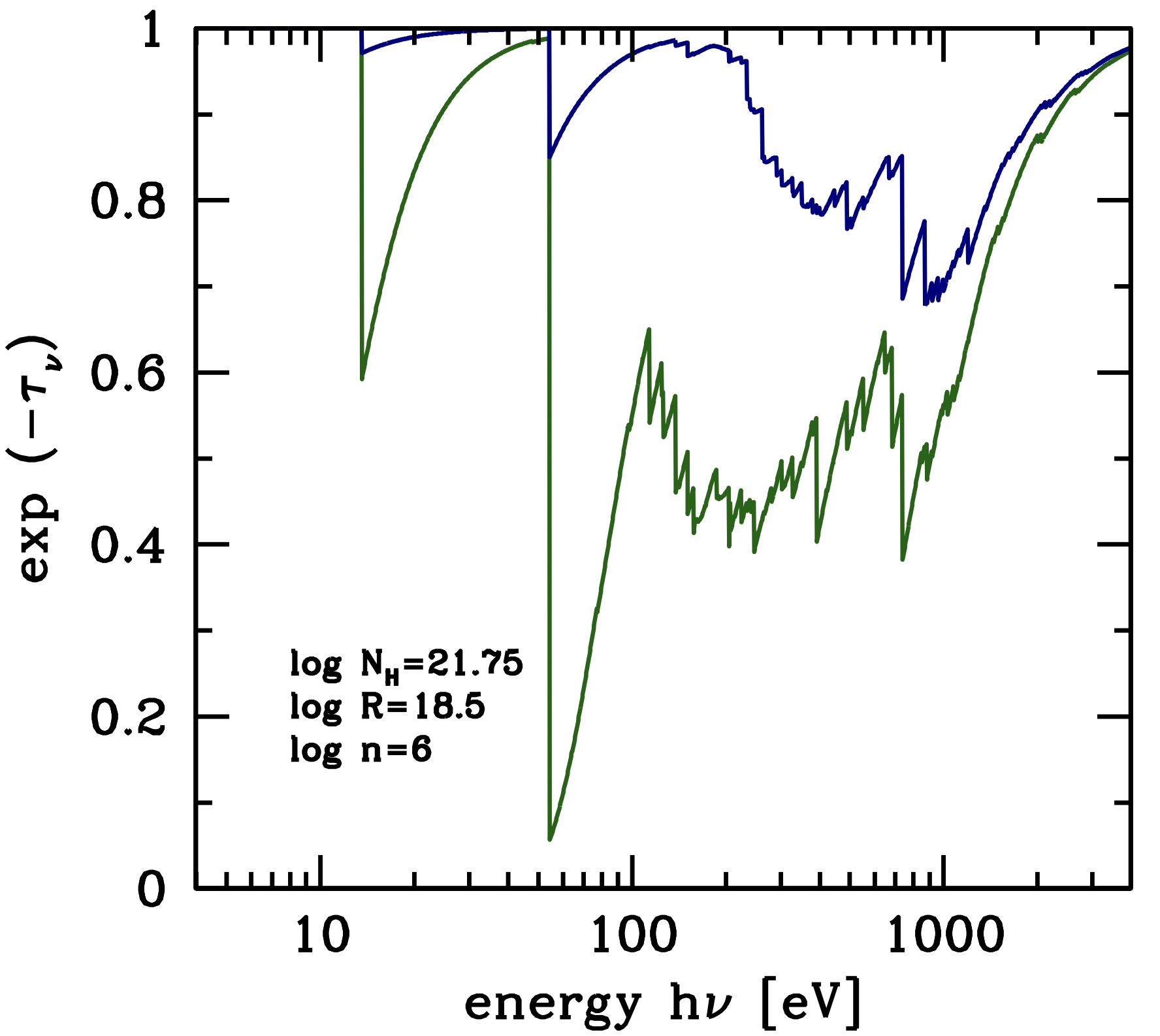}
\caption{Internal, pc-scale AGN absorption in the EUV-soft X-rays. The curves show the transmission through an absorbing medium of column $N_{\rm H}=10^{21.75}\,$ cm$^{-2}$ as a function of energy, obtained using Cloudy photoionization models \citep{Ferland2017}. In these examples, absorbers with density $n=10^{6}\,$cm$^{-3}$ and solar metallicity reprocess the ionizing continuum emitted by an AGN at a distance of $R=10^{18.5}\,$cm. Blue curve: $(M_{\rm UV},\alpha_{\rm EUV})=(-19.5,-1.8)$. Green curve: $(M_{\rm UV},\alpha_{\rm EUV})=(-18.5,-1.4)$. From the emergent filtered photon spectrum, we derive using Equation (\ref{eq:fesc}) the escape fractions $\fescH=0.99$ and $\fescHe=0.93$ for the brighter, steep-spectrum AGN and $\fescH=0.82$ and $\fescHe=0.33$ for the fainter, flat-SED case.
}
\label{fig:CLOUDY}
\end{figure}

Two examples displaying the sensitivity of the EUV transmittance of the absorbing medium to source luminosity (and therefore  ionization parameter) are shown  in Figure \ref{fig:CLOUDY}. Here, we plot the transmission through a solar metallicity column of $N_{\rm H}=10^{21.75}\,$cm$^{-2}$, obtained using Cloudy photoionization models \citep{Ferland2017} for 
different combinations of absolute UV magnitude and EUV spectral slope -- $(M_{\rm UV},\alpha_{\rm EUV})=(-19.5,-1.8)$ (blue curve) and $(M_{\rm UV},\alpha_{\rm EUV})=(-18.5,-1.4)$ (green curve). From the emergent filtered photon spectra, we calculate $\fescH=0.99$ and $\fescHe=0.93$ in the brighter, steeper-spectrum case and $\fescH=0.82$ and $\fescHe=0.33$ for fainter, flatter-spectrum AGNs. Similar transmissions can be estimated for denser, closer-in BLR clouds of comparable columns and ionization parameters, which may also contribute to the modulation of the emerging ionizing spectrum. {In JWST-selected AGNs, the presence of dense, dust-free gas in the nuclear region with large covering factors is also supported by the large equivalent width of the broad component of the H$\alpha$ line \citep{Maiolino2024}.}

We shall assess the possible impact of internal \HeII\ absorption and the resulting shift in the epoch of \HeII\ reionization in Section\,4.

\subsection{Radiative Recombinations}

There are two further sinks of escaping LyC radiation during reionization: hydrogen atoms in the diffuse IGM -- both never ionized before as well as previously photoionized and subsequently recombined -- and the optically thick Lyman-limit systems (LLSs) \citep[e.g.,][]{miralda00,gnedin06,furlanetto09,kaurov15,madau17}. 
The high \HI\ column densities of the LLSs imply an origin in dense environments such as galaxy halos and the circumgalactic medium \citep[e.g.,][]{Kohler2007,Rudie2012, Shen2013}, and the associated photon losses can -- somewhat artificially --  be treated as a reduction in the source term and included in the LyC escape fraction \citep{kaurov15}. Conversely, radiative recombinations in the diffuse IGM 
can be approximately accounted for as follows. We use the case B recombination coefficient (sum over all states except the ground) of hydrogen \citep{Pequignot1991},

\begin{equation}
\alpha_B(T)=10^{-12.366}\,{\rm cm^3\,s^{-1}}\,
\left({T_4
^{-0.6166}\over 1+0.6703\,T_4^{0.53}}\right),
\end{equation}
where $T_4$ is the gas temperature in units of $10^4\,$K. The rationale behind the choice of case B may lie in detailed calculations of the recombination emissivity of the IGM \citep{haardt96,Faucher2009,haardt12}, showing an increasing contribution of reemissions to the photoionization rates near the epoch of 
reionization (when the mean free path of LyC radiation is small and most recombination photons are absorbed before being redshifted below threshold), but there are uncertainties \citep[e.g.,][]{kaurov14}. We therefore fix the temperature of ionized gas to $T_0= 10^{4}\,$K (for consistency with the adopted value of the recombination clumping factor below), noting that gas at this temperature recombines in case B at the same rate of gas at $T = 10^{4.3}\,$K in the oft-used case A situation. Average temperatures higher than $10^4\,$K are typically found for the recombining IGM at $5<z<7$ in hydrodynamical simulations of late reionization \citep[e.g.,][]{Doussot2019,Eide2020,Villasenor2022,Garaldi2022}.

Since the recombination rate is quadratic in density and depends on temperature, its volume average must be computed with the help of cosmological hydrodynamic simulations \citep[e.g.,][]{finlator12,Shull2012,So2014,Jeeson2014,Pawlik2015, kaurov15,Chen2020}. 
Denoting, as is customary, the volume filling factor of ionized hydrogen as $\QH\equiv \langle x_{\rm HII}\rangle$ (here and below the angle brackets denote an average over all space), it is convenient to define a volume-averaged recombination rate per hydrogen atom as $\QH/\bar t_{\rm rec,H}$, where the ``effective" recombination timescale for \HII\ regions in the IGM is \citep{Madau1999}

\begin{equation}
1/\bar t_{\rm rec,H}\equiv (1+\chi)\,\alpha_B(T_0)\,\langle n_{\rm H}\rangle (1+z)^3\,C_R.
\end{equation}
Here, $\langle n_{\rm H} \rangle$ is the mean comoving cosmic density of hydrogen atoms and $\chi\equiv Y/4X$, 
i.e. we have assumed that helium is singly ionized at the same time as hydrogen and doubly ionized only at later times. For the clumping factor $C_R$  that corrects radiative recombinations for density, temperature, and ionization inhomogeneities, we use the results of radiation-hydrodynamic simulations of inhomogeneous reionization by \citet{Finlator2009} and \citet{Chen2020}. A detailed comparison between the volume filling factor model of reionization history (see Section\,4) and the results of the SCORCH reionization simulation project \citep{Chen2020} shows consistency for
\begin{equation}
C_R=\langle x_{\rm HII}\,\rangle C_{\rm HII,10^4\,K}=9.25-7.21\log_{10}(1+z), 
\label{eq:CR}
\end{equation}
where the best-fitting function for the hydrogen clumping factor weighted by ionized volume fraction, $\langle x_{\rm HII}\,\rangle C_{\rm HII,10^4\,K}$,  is from \citet{finlator12}. The above expression yields $C_R=3.2$ at $z=6$, in line with determinations by \citet{pawlik09}, \citet{Shull2012}, \citet{Jeeson2014}, and \citet{Chen2020}. An extrapolation to $z=3$, near the completion of helium reionization, gives $C_R=4.9$, in good agreement with the simulation results on \HeIII\ recombinations by \citet[][]{Jeeson2014}. 
We shall therefore assume that \HII\ and \HeIII\ gas have similar recombination clumping factors, denote the volume filling factor of doubly ionized helium as $\QHe\equiv \langle x_{\rm HeIII}\rangle$, and define the effective recombination timescale for \HeIII\ regions in the IGM as
\begin{equation}
{1/\bar t_{\rm rec,He}}\equiv (1+2\chi)Z\alpha_B(T_0/Z^2)\,\langle n_{\rm H}\rangle C_R,
\end{equation}
where $Z=2$ is the ionic charge.

\begin{figure*}[!ht]
\centering
\includegraphics[width=0.49\hsize]{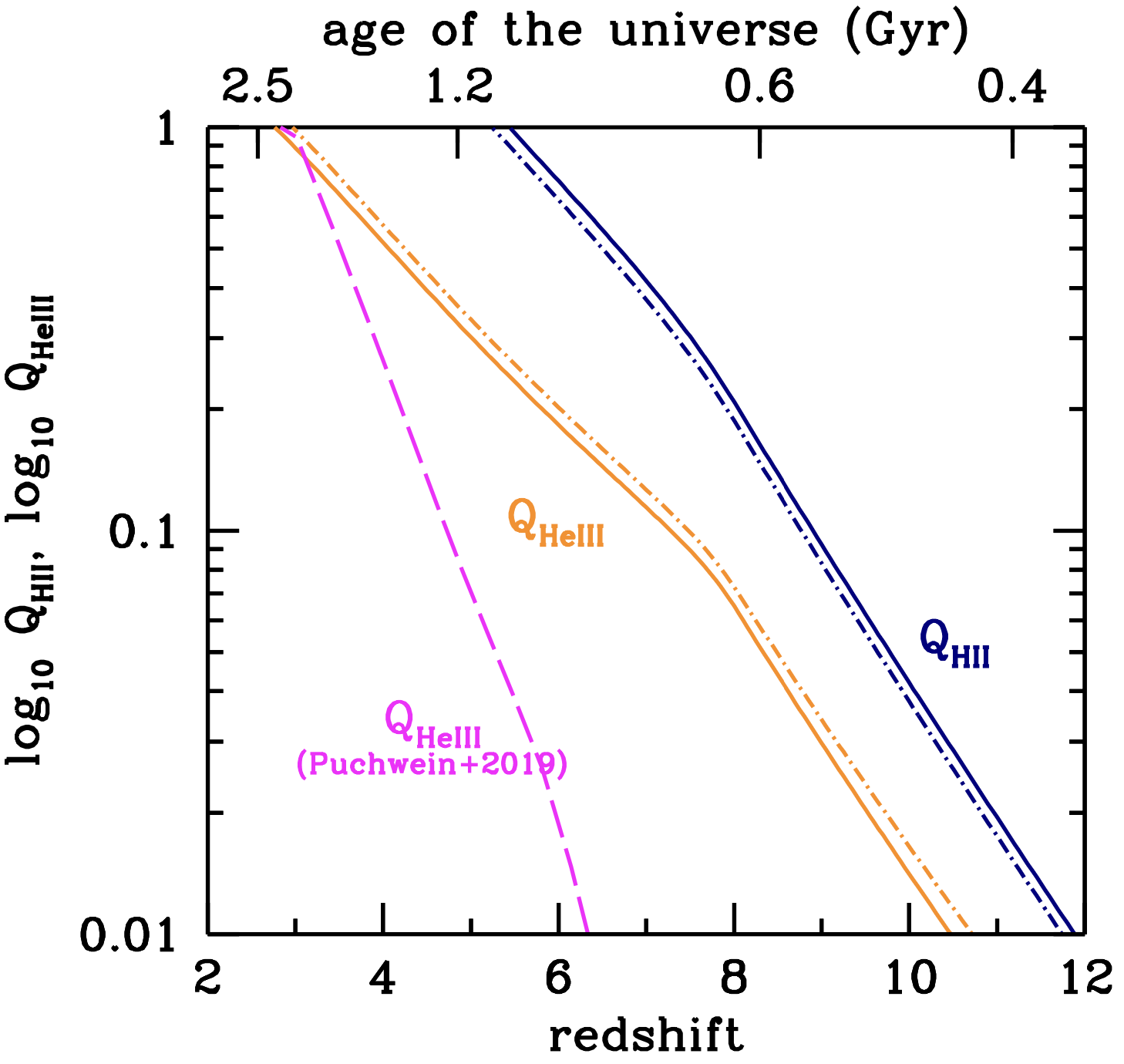}
\includegraphics[width=0.48\hsize]{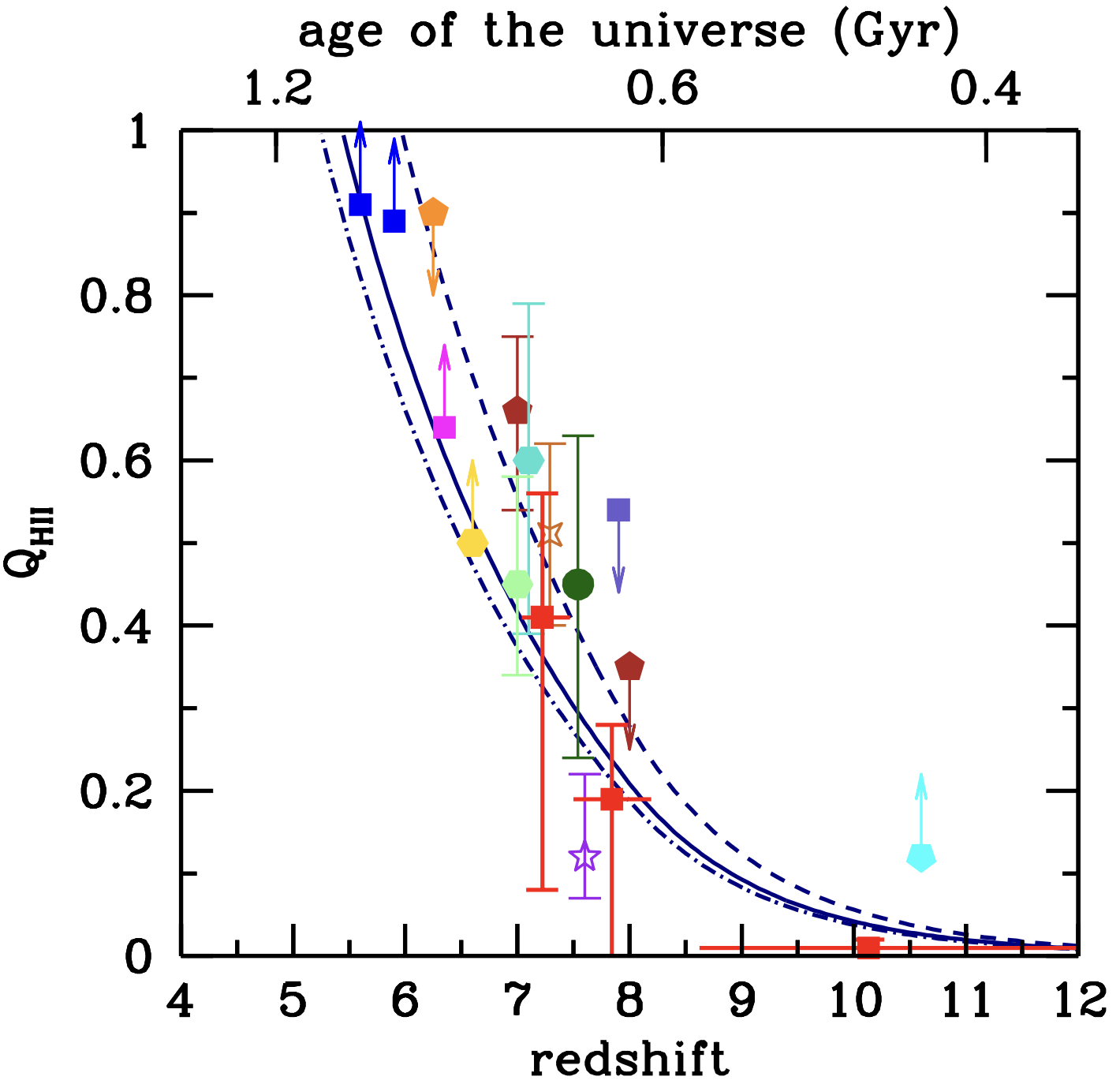}
\caption{Left panel: fiducial reionization histories for our AGN-dominated Models I (solid lines) and II (dotted-dashed lines) (see the main text for details). The evolving \HII\ and \HeIII\  ionized volume fractions are depicted in blue and orange colors, respectively. Hydrogen in the IGM is fully reionized when $Q_{\rm HII}=1$, while helium is fully doubly reionized when $Q_{\rm HeIII}=1$. The dashed curve shows a model of \HeII\ reionization history driven by rare, luminous quasars \citep[the ``fiducial model" of\,][]{puchwein19}.
Right panel: constraints on the ionization state of the IGM from a variety of probes and techniques. The solid, dotted-dashed, and dashed curves denote the ionized volume filling factor $Q_{\rm HII}$ (linear scale) for our Model I, Model II, and ``modified Model I," respectively (see the main text for details). The data points are from  \citet{mcgreer15} (blue squares), \citet{schenker14} (firebrick pentagons), \citet{greig17} (turquoise pentagon), \citet{schroeder13} (orange pentagon), \citet{ouchi10} (gold hexagon), \citet{gallerani08} (magenta square), \citet{banados18} (green dot), \citet{hoag19} (purple star), \citet{bradley22} (brown star), \citet{mason19} (blue square), \citet{Whitler2020} (light-green hexagon), \citet{Nakane2024} (red squares), and \citet{Bruton2023} (cyan pentagon). 
}
\label{fig:poroHHe}
\end{figure*}

\section{Reionization History}

The general idea of a gradual reionization process driven by a steadily increasing UV photon production rate can be modeled analytically by  
the ``reionization equations" \citep{Madau1999}:
\begin{align}
{d\QH \over dt} & ={\dot n_{\rm ion,H}\over  \langle n_{\rm H}\rangle}-{\QH \over \bar t_{\rm rec,H}},\\
{d\QHe\over dt} & ={\dot n_{\rm ion,He}\over \langle n_{\rm He}\rangle}-{\QHe\over \bar t_{\rm rec,He}},
\end{align}
where gas densities are volume-average expressed in comoving units, and we do not explicitly follow the transition from neutral to singly ionized helium, as this occurs nearly simultaneously to and cannot be readily decoupled from the reionization of hydrogen. The above ordinary differential equations assume that the mean free path of UV radiation is always much smaller than the horizon (``local source approximation") and that the absorption of photons above 4 ryd is dominated by \HeII.\footnote{A revised hydrogen reionization equation that explicitly accounts for the presence of optically thick LLSs regulating the mean-free path of LyC  radiation after overlap can be found in \citet{madau17}.}\,
In what follows, we present two models of AGN-driven reionization that yield similar results and fit a number of observational constraints. 

\subsection{Model I: Internal Absorption}

Our ``Model I" is characterized by a flat EUV SED with $\alpha_{\rm EUV}=-1.4$ and by local pc-scale absorption with  $\fescH=0.8$ and $\fescHe=0.3$ (see Figure \ref{fig:CLOUDY}). The parameter $f_{\rm host}$ is set equal to 0.4. Extrapolated to the X-ray band, the assumed SED gives $\alpha_{\rm OX}=-1.35$ for the optical-to-X-ray spectral index needed to join the emissivity at 2500\,\AA\ with that at 2 keV. This value is consistent with the $\langle \alpha_{\rm OX}\rangle=-1.37\pm 0.18$ measured in X-ray selected type 1 AGNs in XMM-COSMOS \citep{Lusso2010}.

The predictions from this model are depicted in Figure \ref{fig:poroHHe} (left panel). Hydrogen reionization is 99\% completed ($\QH=0.99$) by $z=5.45$, while helium is almost fully doubly reionized ($\QHe=0.99$) by $z=2.75$. The midpoints of hydrogen and helium reionization are at $z=6.7$ and $z=4.05$, respectively. 
As shown in the right panel of the figure, the evolving $Q_{\rm HII}$ is broadly consistent with measurements of the changing neutrality of intergalactic hydrogen at $z\gta 6$ from a variety of probes and techniques. Given the large scatter and error bars in these determinations, however, bounds on the hydrogen reionization history are fated to be weak. Perhaps more importantly, the delay in the epoch of \HeII\ reionization resulting from intrinsic absorption and partial leakage at 4 ryd makes an AGN-dominated scenario consistent with observations of the \HeII\ Lyman-$\alpha$ forest, suggesting helium became fully ionized around redshift $z\sim 3$ \citep[e.g.,][and references therein]{Madau1994,Reimers1997,Heap2000,Zheng2004,Dixon2009, worseck16}. Indeed, the extended \HeII\ reionization history envisioned in the left panel of the figure appears to reproduce  observations requiring that the bulk of intergalactic helium was already doubly ionized at $z>3.5$ \citep{worseck16}; these measurements stand in conflict with models of \HeII\ reionization driven by rare, luminous quasars \citep[e.g.,][]{McQuinn2009,puchwein19}.

Our reionization histories  are only illustrative  and should not be taken at face value as they are based on a number of simplifying assumptions and poorly known parameters. For instance, one could push the epochs of complete hydrogen and helium reionization to earlier times by adopting a smaller mean contribution of the host galaxy to the rest-frame UV light. With
$\bar f_{\rm host}=0.2$, Model I would predict $\QH=0.99$ at $z=5.95$ and $\QHe=0.99$ at $z=3.3$,
with midpoint redshifts of 7.15 and 4.6 for hydrogen and helium reionization, respectively. We plot the hydrogen reionization history of this ``modified Model I" in the right panel of Figure \ref{fig:poroHHe} to highlight the sensitivity of 
$\QH(z)$ to uncertainties in the adopted  parameters. 

\subsection{Model II: Steep EUV Spectra}

Uncertainties in the $>1$ ryd AGN SED allow an equally successful AGN-dominated ``Model II," where intrinsic absorption is negligible, but the ionizing continuum is significantly steeper than in Model I. Here, we assume $\fescH=\fescHe=0.9$ and $\alpha_{\rm EUV}=-1.9$. This power-law slope is consistent with the $\langle \alpha_{\rm EUV}\rangle =-1.70\pm 0.61$ and 
$\langle \alpha_{\rm EUV}\rangle =-1.96\pm 0.15$
measured by \citet{Lusso2015} and \citet{Zheng1997}, respectively. Extrapolated to the X-ray band, this SED is characterized by a steeper $\alpha_{\rm OX}=-1.80$ for the optical-to-X-ray spectral slope between  2500\,\AA\ and 2 keV. This spectral index matches the value $\alpha_{\rm OX}=-1.80\pm 0.02$ measured in the most luminous optically selected quasars at $z=1.5-4.5$ \citep{Just2007}. As before, we fix $f_{\rm host}=0.4$.

The predictions from this model are depicted in Figure \ref{fig:poroHHe} and are very similar -- by design -- to those of Model I. Hydrogen reionization now completes a bit later than in Model I, $\QH=0.99$ at $z=5.25$, while helium is doubly reionized a bit earlier,  $\QHe=0.99$ at $z=2.95$. The midpoints of hydrogen and helium reionization in Model II are at $z=6.5$ and $z=4.25$, respectively. A late, $z\lta 5.3$, end of hydrogen reionization may explain the presence of excess scatter in the Lyman-$\alpha$ forest at $z\sim 5.5$, together with the existence of sporadic extended opaque Gunn-Peterson troughs \citep{Bosman2022}. In both Models I and II, the contribution of $z\gta 10$ AGNs to reionization is negligible.

The above scenarios highlight the constraints that must be satisfied by a successful AGN-dominated framework and set the context for a reexamination of the role played by AGNs in the reionization of  hydrogen and helium in the IGM.

\begin{figure}[!ht]
\centering
\includegraphics[width=0.9\hsize]{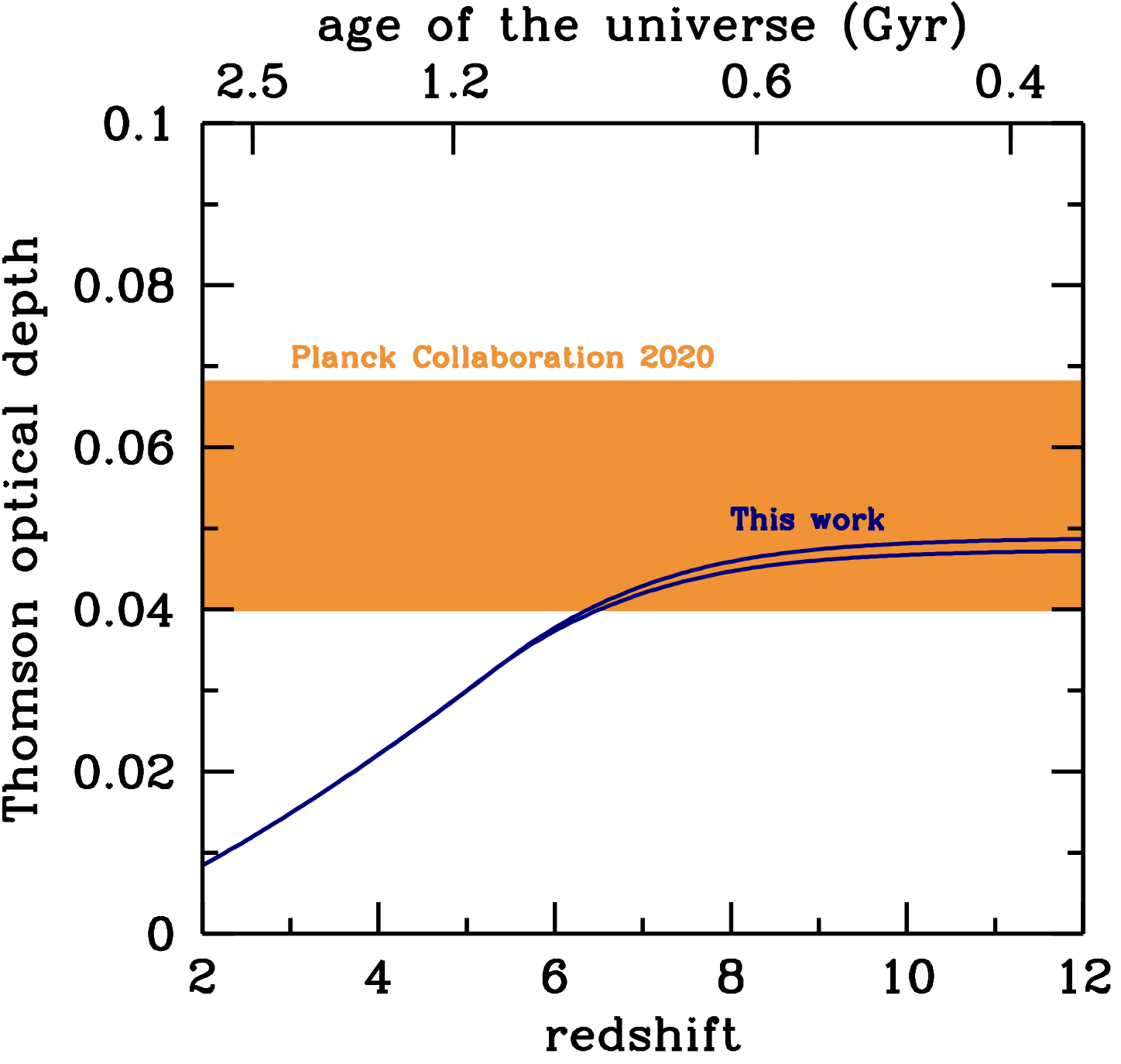}
\includegraphics[width=0.9\hsize]{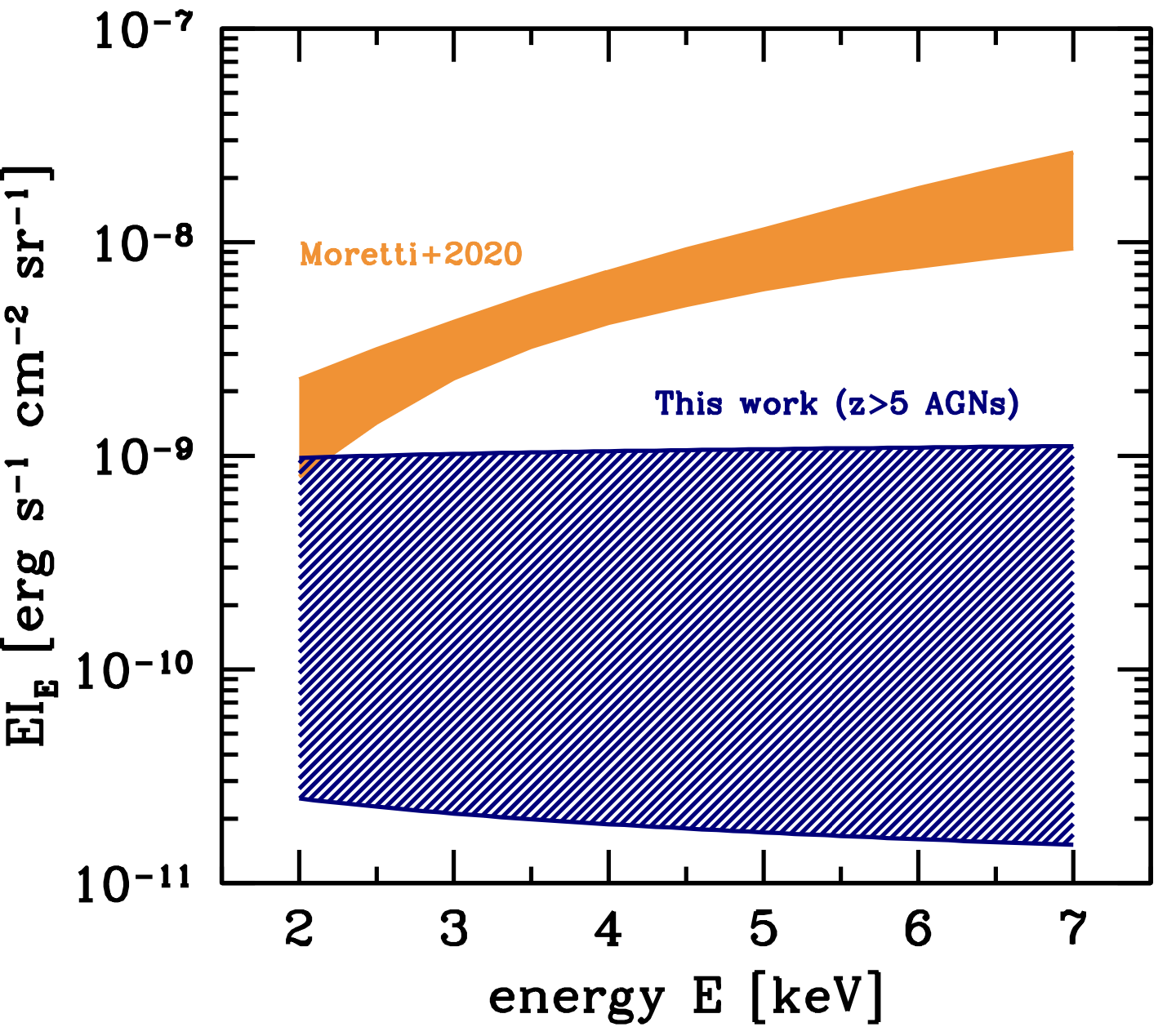}
\caption{Top panel: optical depth to reionization as a function of maximum integration redshift $z$, for our AGN-dominated Model I (upper blue
curve) and Model II (lower blue curve). The
\citet{Planck2020} opacity constraint, $\tau = 0.054 \pm 0.014$ (2$\sigma$),  is shown as the orange area.
Bottom panel: constraints from the unresolved $2-7$ keV XRB. The blue upper envelope marks the predictions of an
AGN-dominated ``high-background" case with $\alpha_{\rm OX}=-1.35$, $\alpha_X=-0.9$, and $R_{\rm II}=2$ (see the main text for details).
The lower envelope depicts a ``low-background" case with $\alpha_{\rm OX}=-1.8$, $\alpha_X=-1.4$, and $R_{\rm II}=2$. The orange shading shows the spectrum of the unresolved XRB from \citet{Moretti2012}.
}
\label{fig:tauXRB}
\end{figure}

\subsection{Integrated Optical Depth to Reionization}

Free electrons generated during reionization scatter and partially damp cosmic microwave background (CMB) anisotropies. CMB observations  therefore provide limits on the epoch and duration of the reionization process that are highly complementary to those obtained from other probes. We calculate the integrated electron scattering optical depth in our models as
\begin{equation}
\tau(z)=c\sigma_T \langle n_{\rm H}\rangle \int_0^{z}{(1+z')^2dz'\over H}{\cal Q}(z'),
\label{eq:taues}
\end{equation}
where $c$ is the speed of light, $\sigma_T$ the Thomson cross-section, $H(z)$ is the Hubble parameter, 
${\cal Q}\equiv \QH (1+\chi)+ \chi\QHe$, 
and we have assumed $Q_{\rm HeII}=\QH-\QHe$. As shown in Figure 
\ref{fig:tauXRB}, Models I and II produce Thomson scattering optical depths to reionization of $\tau=0.049$ and $\tau=0.047$, respectively, both consistent with the value reported by the Planck collaboration,
$\tau= 0.054 \pm 0.007$ at the $1\sigma$ level \citep{Planck2020}.\footnote{Note that  in Equation (\ref{eq:taues}) the correct calculation of the mean electron density should use a  mass-weighted ionization fraction. Our results may therefore be slightly biased compared to radiation-hydrodynamic simulations \citep[][]{Chen2020}.}\,

\subsection{Constraints from the X-Ray Background}

Despite nondetections in the X-ray, as may be expected if they are below current survey limits, the population of broad-line AGNs observed in recent JWST surveys may contribute to the observed X-ray background (XRB). Indeed, it has been recently argued by \citet{Padmanabhan2023} that the inferred  AGN UV emissivities at $z>5$ imply an unresolved XRB that is about 1 order of magnitude higher than constrained by current experiments. Here, we show that this is actually not the case.\footnote{We note that Equation (3) in \citet{Padmanabhan2023}, defining the ``Olbers' integral" for the specific intensity of the XRB is actually incorrect. In their integrand expression there is a missing factor of $(1+z)$ in the numerator associated with the reduction of the bandwith.}\, The expected contribution at energy $E$ from the AGN
population above redshift $z_X$ can be expressed as \citep{Haardt2015,madau15}
\begin{equation}
I_E={c\over 4\pi}\,\left({E\over 2\,{\rm keV}}\right)^{\alpha_X}\,\int_{z_X}^\infty {dz\over H}\,\epsilon_{2}(z)\,(1+z)^{\alpha_X-1},
\label{eq:XRB}
\end{equation}
where the specific comoving emissivity at 2 keV, $\epsilon_{2}(z)$, is related to the AGN emissivity at the  Lyman limit by
\begin{equation}
\begin{aligned}
\epsilon_2(z)= & (1-\bar f_{\rm host})\,\epsilon_{912}(z)
\,R_{\rm II}\\
& \times \left({912\over 2500}\right)^{\alpha_{\rm FUV}}\left({2\,{\rm keV}\over 0.005\,{\rm keV}}\right)^{\alpha_{\rm OX}}.
\label{eq:epsi2}
\end{aligned}
\end{equation} 
Here, $\alpha_X$ is the spectral index (corresponding to a photon index $\Gamma=\alpha_X-1$) at rest-frame energies above 2 keV.
At $z>5$, the $2-7$ keV band probes $12-40$ keV radiation in the rest frame, which should basically be obscuration independent. For this reason we include in Equation (\ref{eq:epsi2}) a correction factor, $R_{\rm II}$, which accounts for the possible contribution of UV-obscured (``type-2"), narrow-line AGNs at $z\ge z_X$ to the observed XRB. A type 2 AGN/galaxy fraction of about 20\% at $z\sim 4-6$ has been recently estimated by \citet{Scholtz2023}, and we therefore assume $R_{\rm II}=2$ below. 

We use $\alpha_{\rm FUV}=-0.61$ and 
$\bar f_{\rm host}=0.4$ as before, and integrate Equation (\ref{eq:XRB}) for $z>z_X=5$ for two different scenarios to bracket the uncertainties: (1) a ``high-background" case with $\alpha_{\rm OX}=-1.35$ (as in Model I) and $\alpha_X=-0.9$ (the canonical value for X-ray selected AGNs, see, e.g., \citealt{Merloni2014}); and (2) a ``low-background" case with $\alpha_{\rm OX}=-1.8$ (as in Model II) and  $\alpha_X=-1.4$, the steep value that is characteristic of luminous, $z>6$ quasars \citep{Zappacosta2023} {and of close-to-Eddington sources \citep{Tortosa2023}}. As shown in Figure \ref{fig:tauXRB}, even for the high-background extreme parameters we obtain $EI_E\simeq 10^{-9}$ erg s$^{-1}$ cm$^{-2}$ sr$^{-1}$ at 2 keV, a contribution that would not saturate the unresolved XRB measured by \citet{Moretti2012}. The low-background case lies well below the \citet{Moretti2012} determinations. 

These results show that $z>5$ AGNs can reionize the Universe without overproducing the unresolved XRB  even if their properties (i.e., fraction of obscured objects, optical-to-X-ray spectral indices) are similar to those of their lower-redshift counterparts, a conclusion analogous to that reached by \citet{madau15}. 

\section{Summary and Conclusions}

Deep spectroscopic surveys with JWST are opening a new parameter space for low-luminosity AGN activity at high redshift, leading to a shift in our understanding of the formation and growth of massive black holes in the early Universe. The large LyC production efficiency and leakiness into the IGM typical of UV-selected AGNs at lower redshifts, together the high AGN number fractions reported, have led us to reassess a scenario where type 1 AGNs are the main drivers of the cosmic hydrogen/helium reionization process. Our approach moves the spotlight away from nonactive faint star-forming galaxies if, as we surmise, LyC leakage is either ``on or off'' owing to AGN activity. It is based on the assumptions, grounded in recent observations, that (1) the fraction of type 1 AGNs among galaxies is around $10–15\%$; (2) the mean escape fraction of hydrogen LyC radiation is high, $\gta 80$\%, in AGN hosts  and is negligible otherwise; (3) internal absorption at 4 ryd or a steep ionizing EUV spectrum delays the epoch of \HeII\ full reionization and makes an AGN-dominated scenario consistent with observations of the \HeII\  Lyman-$\alpha$ forest
\citep[cf.][]{madau15,daloisio17,Mitra2018,Garaldi2019}. 
In our fiducial models, (1) hydrogen reionization is 99\% completed by redshift $z\simeq 5.3-5.5$, and reaches its midpoint at $z\simeq 6.5-6.7$; (2) the integrated Thomson scattering optical depth to reionization is $\simeq 0.05$, consistent with constraints from CMB anisotropy data; (3) partially absorbing gas layers on parsec scales moderate the leakage of $> 54.4$ eV helium-ionizing photons ($\fescHe\simeq 30\%$), or the intrinsic EUV spectra are steep ($\alpha_{\rm EUV}\simeq -1.9$). The emerging AGN emissivity doubly reionizes helium by redshift $z\simeq 2.8-3.0$; (4) the abundant AGN population detected by JWST does not violate constraints on the unresolved XRB, as the Olbers' integral for the background intensity from $z>5$ sources lies $>1$ order of magnitude below the observations in the case of a steep  X-ray AGN SED; and (5) the contribution of $z\gta 10$ AGNs to hydrogen reionization is minor.

Obviously, given the uncertainties, these numbers are only meant to be indicative and should be viewed in the context of a reexamination of the role played by AGNs in the ionizing photon budget of the $z>5$ IGM. We note here that the scale of the H-ionizing emissivity is actually set by the product $f_{\rm AGN}\,\fescH\,(1-\bar f_{\rm host})$. The quoted values of these parameters are therefore not critical since different combinations can generate similar reionization histories. {However, our fiducial models typically require $f_{\rm AGN}\,\fescH\,(1-\bar f_{\rm host})\gta 0.08$. A fraction of AGNs significantly below 10\%  would then not suffice at making accretion onto massive black holes the sole driver of cosmic reionization and would require a contribution from nonactive galaxies. As an example, let us write the total normalization of a mixed model as $\sim f_{\rm AGN}\,\fescH({\rm AGN})\,(1-\bar f_{\rm host}) + (1-f_{\rm AGN})\,\fescH({\rm GAL})\,D_{912}$, where $D_{912}$ is the intrinsic discontinuity at 1 ryd of massive stars. Assuming, for illustration purposes, 
$\fescH({\rm AGN})=1.0$, $\bar f_{\rm host}=0$, and $D_{912}=1/3$, 
a scenario with $f_{\rm AGN}\sim 0.05$ and $\fescH({\rm GAL})\sim 0.1$
would produce a hydrogen reionization history similar to those depicted in Figure \ref{fig:poroHHe}.
By the same token, the contribution of JWST-discovered AGNs to reionization would be small compared to that of nonactive galaxies if the X-ray weakness of the former was the consequence of heavy X-ray absorption by Compton thick, cold nuclear material \citep{Maiolino2024}.}

Taken at face value, our results should promote further testing of AGN-dominated reionization scenarios against new observations. In particular, internal absorption and/or steep EUV AGN spectra are required to delay the epoch of \HeII\ complete reionization to redshift $z\sim 3$, make an AGN-dominated scenario consistent with the observed slow evolution of the \HeII\ Lyman-$\alpha$ opacity \citep{worseck16} -- in our models the \HeIII\ volume fraction is already $25-30$\% when hydrogen becomes fully reionized -- and remove tensions with constraints on the thermal history of the IGM \citep[e.g.,][]{daloisio17,Villasenor2022}. Our estimates are based on early JWST observations that do not yet depict a clear picture of the type 1 AGN/galaxy fraction {and internal absorption properties} as a function of redshift. If proved true, however, these models may require  a complete revision to the standard view that nonactive galaxies dominate the high-redshift hydrogen-ionizing background.

\bigskip
\section*{Acknowledgements}
Support for this work was provided by NASA through grant 80NSSC21K027 (PM). A.G. and E.G. acknowledge the support of the INAF GO/GTO grant 2023 ``Finding the Brightest Cosmic Beacons in the Universe with QUBRICS'' (PI Grazian). A.G. acknowledges the support of the INAF Mini grant 2022 ``Learning Machine-Learning techniques to dig up high-$z$ AGNs in the Rubin-LSST Survey'' and the PRIN 2022 project 2022ZSL4BL INSIGHT, funded by the European Union--NextGenerationEU RFF M4C2 1.1.

\bibliographystyle{apj}
\bibliography{paper}

\label{lastpage}
\end{document}